\documentclass{sarticle}
\input{preamble.tex}
\hypersetup{
    pdftitle={One-dimensional quantum gravity and the Schwarzian theory},
    pdfauthor={D. Anninos, D. M. Hofman, S. Vitouladitis}
}

\usepackage{nicefrac}
\newcommand{\sv}{\(\ddagger\)}
\renewcommand{\dh}{\(\dagger\)}
\newcommand{\da}{\(\ast\)}
\newcommand{\uva}{2}
\newcommand{\kcl}{1}

\title{One-dimensional Quantum Gravity and the Schwarzian theory}
\author[\kcl,\da]{Dionysios Anninos}
\emailAdd[\da]{dionysios.anninos@kcl.ac.uk}

\author[\uva,\dh]{Diego M. Hofman}
\emailAdd[\dh]{d.m.hofman@uva.nl}

\author[\uva,\sv]{Stathis Vitouladitis}
\emailAdd[\sv]{e.vitouladitis@uva.nl}

\affiliation[\kcl]{Department of Mathematics, King’s College London, the Strand, London WC2R 2LS, U.K.}
\affiliation[\uva]{Institute for Theoretical Physics, University of Amsterdam, Science Park 904, Postbus 94485, 1090 GL Amsterdam, the Netherlands}

\abstract{We develop a model of one-dimensional (Conformal) Quantum Gravity. By discussing the connection between Goldstone and Gauge theories, we establish that this model effectively computes the partition function of the Schwarzian theory where the \(\SL(2,\mathbb{R})\) symmetry is realized on the base space. The computation is straightforward, involves a local quantum measure and does not rely on localization arguments. Non-localities in the model are exclusively related to the value of fixed gauge invariant moduli. Furthermore, we study the properties of these models when all degrees of freedom are allowed to fluctuate. We discuss the UV finiteness properties of these systems and the emergence of a Planck's length.
}

\newcommand{\be}{\begin{equation}}
\newcommand{\ee}{\end{equation}}
\newcommand{\bea}{\begin{eqnarray}}
\newcommand{\eea}{\end{eqnarray}}

\newcommand{\lp}{\ell_\t{P}}
\newcommand{\ads}{\t{AdS}}
\newcommand{\ds}{\t{dS}}
\newcommand{\cft}{\t{CFT}}
\newcommand{\diffu}{\frac{\diff\,}{\U(1)}}
\newcommand{\sch}[2]{\qty{#1,#2}}
\newcommand{\diffs}{\diff}

\begin{document}

\maketitle

\newpage

\section{Introduction}

Over the last two decades, the \(\ads/\cft\) correspondence has produced (or directed interest towards) a large number of boundary theories with extraordinary properties, including supersymmetry and integrability. These are typically CFTs or, more generally, UV complete Quantum Field Theories (QFT). If the bulk description in terms of String Theory in \(\ads\) space is self sufficient, then so is the boundary by necessity. More recently, examples of holography in near-\(\ads_2\) spacetimes have been understood \cite{Maldacena:2016upp,Jensen:2016pah}. In this case we imagine cutting off a part of the \(\ads_2\) disk and formulating a dual theory on this boundary. This way, we are agnostic about the way in which this geometry is asymptotically completed. On the boundary side, these setups give rise to the so-called Schwarzian theory and deformations thereof. These are Goldstone effective theories related to the \(\SL(2,\mathbb{R})\) symmetry group  \cite{Kitaev:2017awl,Maldacena:2016hyu} which are by definition the universal low energy sector of unspecified UV complete quantum mechanical theories. A popular formulation for this completion is given by the SYK model \cite{Sachdev:1992fk,Kitaev:2017awl,Maldacena:2016hyu}, which is not a standard QFT as the coupling constants of the model need to be drawn from a probabilistic distribution.

These Goldstone theories have attracted significant interest over the last few years, both from the point of view of gravitational bulk physics and the boundary quantum chaos community.  While much is known about these theories and their partition function, which admits an exact computation \cite{Stanford:2017thb,Mertens:2017mtv,Belokurov:2017eit}, the Schwarzian theory is still rather complicated in its current formulation. The main reason that this is the case is that the quantum measure associated to the model is not local if \(\SL(2,\mathbb{R})\) symmetry is to be explicitly preserved.  Progress toward understanding the theory completely requires a partial gauge fixing of the degrees of freedom.

In this work we present a formulation of this family of theories in terms of theories of (Conformal) Quantum Gravity in one dimension. In this setup, the quantum measures are fully local and explicitly given in covariant language. The non-localities manifest themselves through the observables that must be computed. We discuss the general theory of how to use Gauge theories to describe Goldstone theories and then apply it concretely to the problem at hand.

The theories of Quantum Gravity suggested by this process are the closest one-dimensional analog to the worldsheet theories of String Theory. They present unexpected properties and have potentially  interesting observables. In particular they provide an interesting example of a UV complete theory of Quantum Gravity where the Planck length is emergent. We explain this feature in detail and give arguments why any UV complete theory of Quantum Gravity must present non-local features of the type seen in the Schwarzian theory. In particular, the Hilbert space interpretation of the matter sector of the Quantum Gravity theory must necessarily break down at the Planck scale and new \emph{ghost} states must appear at this scale making the effective dimension of the Hilbert space zero. We give a detailed account of this physics in the body of this work.

This note presents a first step in understanding theories of one-dimensional Quantum Gravity as boundary models that can exhibit, also gravitating, dual descriptions. This is an important part of the puzzle in understanding holography for  \(\ds\) spacetimes and bulk observers  \cite{Anninos:2011af,Anninos:2018svg,Anninos:2017hhn}.

\section{Goldstone and Gauge theories}
\label{sec:GG}

Traditionally, Goldstone theories are discussed in the context of local Quantum Field Theory. This is quite expected as they appear in the Infrared as effective descriptions of physical systems. This truism gets slightly challenged when one considers Goldstone theories with non-compact symmetries. Here one encounters an obvious obstacle: the partition function of theories with non-compact symmetries is always formally divergent.\footnote{Let us consider for simplicity the case of target space symmetries.} This leads to the following unfamiliar but correct conclusion: physical systems cannot have global non-compact symmetries. They either get broken (\emph{put the system in a box}), compactified (\emph{put the system on a circle}) or gauged (\emph{ignore properly defined infinite group volume factors}). This last option is the typical choice for Goldstone systems. We define the partition function as:
\begin{align}
  \parti_\t{Goldstone} \coloneqq \int \frac{\DD{\varphi}}{\vol(G)} \ \ex{-\act[\varphi]}~,
\end{align}
where \(\varphi\) is a collection of Goldstones, \(G\) is the non-compact symmetry group, and \(\DD{\varphi}\) is an invariant quantum measure. Notice that dividing by the group volume is a non-local procedure. Furthermore, the invariant a quantum measure for \(\varphi\) can also present mild non-localities. (The Schwarzian theory is one recent example \cite{Stanford:2017thb}.)

Fortunately, there is a remedy for this unpleasant situation. Gauge theories are precisely engineered to perform the above procedure in a manifestly local way.  {In this note we focus on $(0+1)$-dimensional systems.  Although Goldstone theories in less than three spacetime dimensions may sound unusual, we follow the conventional abuse of language in the literature. These are theories that remain gapless even though they might not be protected by the phenomenon of symmetry breaking due to large IR fluctuations in the vacuum state.\footnote{A two dimensional example is a free boson below the KT transition. Symmetry breaking might also arise in the large $N$ limit in low dimensions \cite{Witten:1978qu}.} In the presence of non-compact symmetries, the vacuum state ends up being non-normalizable and the problem is even more severe, as mentioned above. Then gauging is necessary. Given that gauge fields are non-dynamical in $(0+1)$-dimensions,\footnote{Let us comment briefly on higher dimensional examples. In $(1+1)$-dimensions there are no propagating degrees of freedom for the gauge field. Nonetheless, the gauge singlet constraint is slightly richer and involves the gauge field in a more elaborate way \cite{tHooft:1974pnl}. In higher than two spacetime dimensions,  new local dynamical degrees of freedom are introduced upon gauging the theory and these must be appropriately decoupled if we are to isolate the original Goldstone theory in the Infrared. This can happen naturally if the gauge theory is IR free. If this is not the case (as in non-abelian gauge theories or gauged WZW models \cite{Polyakov:1984et,Gawedzki:1991yu,Armoni:2000uw}) then local degrees of freedom need to be removed explicitly and the construction becomes less useful. In general, however, very little is known about the IR properties of non-compact gauge theories.  Alternatively, we might be able to couple the Goldstone theory to a topological gauge field.}  the effect of gauging the original theory is to simply impose the gauge singlet constraint. In terms of the path integral, the only thing we need to worry about is to exclude global gauge invariant degrees of freedom in the form of moduli that might appear in the gauge theory formulation. We will discuss this in great detail in the following sections.}

With this in mind, we define the gauged $(0+1)$-dimensional theory to be
\begin{align}
  \parti_\t{Gauge} \coloneqq \int \frac{\DD{\varphi} \DD{a}}{\vol(\cG)} \ \ex{-\act[\varphi, a]}~,
\end{align}
where we have introduced a local gauge field \(a\). In the above expression the volume factor \(\vol(\cG)\) represents, this time, the volume of the gauge group \(\cG\). This is a local quantity (i.e. a volume of the global group per point in the base space). The measure can always be written in a local form as there is no gauge fixing procedure implied.\footnote{This is, of course, up to anomalies which will not concern us in our examples below.}

How do we go back to the previous description? The standard procedure consists in gauge fixing \(a\) and cancelling the volume factor in the denominator with the redundant functional integration on top (e.g. by using Faddeev--Popov determinants).  The only thing interesting that remains is the integral over moduli of the gauge field, which cannot be fixed away and, still dividing by the zero modes associated with the gauge transformations that leave the gauge field invariant. These are, by definition, the original global symmetries, provided there are no global obstructions. The volume generated by the zero modes can, in principle, depend on the moduli. This generates a measure \(w(\mu)\) in moduli space. Picking a representative gauge fixed field \(a[\mu]\) as function of moduli \(\mu\), the path integral becomes:
\begin{align}
  \parti_\t{Gauge} = \int \dd{\mu} \, w(\mu) \int \frac{\DD{\varphi} }{\vol(G)}  \ \ex{-\act[\varphi, a[\mu]]}~.
\end{align}
From the above expression we see immediately the connection to our Goldstone theory. Define the origin of moduli space \(\mu_0\) to satisfy \(a[\mu_0]=0\). We obtain the following equality:
\begin{align}
  \ev{\delta_w[\mu-\mu_0] }_\t{Gauge} =  \int \frac{\DD{\varphi} }{\vol(G)} \ \ex{-\act[\varphi]}  = \parti_\t{Goldstone}~,
\end{align}
where we have chosen to insert a delta function with respect to the measure \(w(\mu)\). More generally, one can define the Goldstone theory partition function in a general background \(a[\mu]\) as:
\begin{align}
  \ev{\delta_w[\mu -\bar \mu] }_\t{Gauge} =  \int \frac{\DD{\varphi} }{\vol(G)} \ \ex{-\act[\varphi, a[\bar \mu]]}  \eqqcolon \parti_\t{Goldstone}[\bar \mu]~.
\end{align}

\section{A simple example: one-dimensional non-compact \(\U(1)\)}
\label{sec:u1}

Let us work out a simple example of the formalism above. Consider a one-dimensional non-compact \(\U(1)\) Goldstone theory on a circle base space of  unit size parameterized by \(\tau \sim \tau+1\). The action is given by
\begin{align}
  \act[\varphi, a]= \frac{1}{g^2} \int_0^1 \dd{\tau} \, \qty(\partial_\tau \varphi -a)^2~.
\end{align}
The associated measure is local and flat and \(\varphi\) takes values in \(\mathbb{R}\). The associated gauge transformations are:
\begin{align}
  \varphi \mapsto \varphi + \omega~, \qquad a \mapsto a + \partial_\tau \omega~.
\end{align}
This simple example has one associated gauge-invariant modulus, \(\mu = \int_0^1 \dd{\tau} a\), and one zero mode \(\omega = \t{constant}\). The partition function can then be easily computed as:
\begin{align}
  \parti^{\U(1)}_\t{Goldstone}[\bar \mu] = \int \frac{\DD{\varphi} \DD{a}}{\vol(\cG)} \ \delta\qty[ \int_0^1 \dd{\tau}\, a - \bar \mu] \ex{-\act[\varphi, a]}~.
\end{align}
Gauge-fixing \(a\) in the above action will produce the expected definition of the Goldstone theory. While this procedure does not produce any unpleasant features in this case and the path integral can be easily computed, we choose instead to use the gauge freedom to fix \(\varphi=0\). Notice that this leaves no zero mode behind as all degrees of freedom in \(\omega\) are used in the gauge fixing. This simplifies the computation and gives:
\begin{align}
  \parti^{\U(1)}_\t{Goldstone}[\bar \mu] = \int  \DD{a}\ \delta\qty[ \int_0^1 \dd{\tau} a - \bar \mu] \exp(-\frac{1}{g^2} \int_0^1 \dd{\tau} \, a^2)~.
\end{align}
We separate the modulus and perform the integral over all fluctuating modes of \(a\) by usual zeta-functional regularized determinants and obtain:\footnote{In this note we discard numerical overall factors in partition functions. A natural way to normalize them is to associate them to a Hilbert space trace.}
\begin{align}
  \parti^{\U(1)}_\t{Goldstone}[\bar \mu] = \frac{1}{g}\exp(-\frac{\bar \mu^2}{g^2})~.
\end{align}
This is the correct final result. Notice that the path integral above is actually one-loop exact. One could view the \(g^{-1}\) factor as coming from the one-loop determinant corrected for the zero modes and the exponential as a non-perturbative contribution coming from the background solution. If one were to integrate over the modulus one would reproduce the trivial result expected in the Gauge theory:
\begin{align}
  \parti^{\U(1)}_\t{Gauge} =1~,
\end{align}where the measure over moduli space is given by \(\dd{\mu} w(\mu) =  \dd{\mu}\) and \(\mu \in \mathbb{R}\).

\section{The Schwarzian Theory}
\label{sec:s}

The Schwarzian theory is defined to be the Goldstone theory associated to the set of \(\SL(2,\mathbb{R})\) transformations mapping the circle to itself. As such, we consider both base and target spaces to be circles of size \(\beta\). Considering coordinates on both circles in \([0,1]\) we can write the action as:
\begin{align}\label{schact}
  \act_\alpha[\phi] \coloneqq - \ell \int_0^1 \frac{\dd{\tau}}{\beta} \qty[ \sch{\phi}{\tau} - 2 \pi^2 \alpha\, \qty(\partial_\tau \phi)^2]   + \lambda \,\beta  \int_0^1 \dd{\tau} ~,
\end{align}
where
\begin{align}
  \sch{\phi}{\tau} = \partial_\tau^2 \log \partial_\tau \phi - \frac{1}{2} \qty(\partial_\tau \log \partial_\tau \phi)^2~.
\end{align}The action above presents a target space \(\SL(2,\mathbb{R})\) symmetry
\begin{align*}
  \tan(\pi \qty(\phi-\half)) \mapsto \frac{a\, \tan(\pi \qty(\phi-\half)) + b}{c\, \tan(\pi \qty(\phi-\half)) + d},
\end{align*} with \(a\,d-b\,c =1\) for the preferred value \(\alpha=-1\). We consider this value to define the Schwarzian theory.  For values \(\alpha>-1\), the action above is only symmetric under the \(\U(1)\) transformations that act as translations on the circle.\footnote{Examples with \(\alpha \neq -1\) include \cite{Anninos:2018svg,Yoon:2019cql,Witten:2020wvy,Maxfield:2020ale}} We refer to this case as the \(\diffu\) theory. The dimensionful parameter \(\lambda\) is allowed by the symmetry and we will shortly interpret it as a form of cosmological constant. Lastly, \(\ell\) is a UV scale analogous to the stiffness is general models of symmetry breaking.

Let us first concentrate on the \(\alpha=-1\) case. In order to compute this path integral we need to define an \(\SL(2,\mathbb{R})\) measure and integrate over monotonic maps \(\partial_\tau \phi >0\). This results in a non-local measure and the need to subtract explicitly the divergences associated to the \(\SL(2,\mathbb{R})\) zero modes, as discussed in the previous section. This was done explicitly in \cite{Stanford:2017thb,Mertens:2017mtv,Bagrets:2017pwq} by localization and Hamiltonian methods. The result is:
\begin{align}\label{stanwit}
  \parti_\t{Schwarzian} =\int \frac{\DD{\phi}}{\vol(\SL(2,\mathbb{R}))}\ \ex{-\act_{-1}[\phi]} = \qty(\frac{\ell}{\beta})^{\frac{3}{2}}\ \exp(\frac{2 \pi^2 \ell}{\beta} -\lambda \beta)~.
\end{align}
Motivated by the relation between Goldstone and Gauge theories explained in the previous section,  in what follows we will present a theory of Conformal Quantum Gravity that can reproduce the result above. The associated quantum measures will be standard and local and the evaluation of the path integral will be straightforward, illuminating the highly technical previously known procedures. Non-localities will be hiding in the choice of integration region. The resulting theory is of interest beyond its application to the Schwarzian theory discussed here. We will comment on this in section \ref{sec:cqg}.

The situation is similar for the \(\diffu\) theory. The main difference is that the number of zero modes changes giving a different overall power of \(\beta\). Here we quote the result \cite{Saad:2019lba} as
\begin{align}\label{u1sch}
  \parti_{\diffu} =\int \frac{\DD{\phi}}{\vol(\U(1))}\ \ex{-\act_{\alpha}[\phi]} = \qty(\frac{\ell}{\beta})^{\frac{1}{2}}\ \exp(-\frac{2 \pi^2 \ell \alpha}{\beta} -\lambda \beta)~.
\end{align}
This theory, in turn, admits a description in terms of a more familiar one-dimensional Quantum Gravity.

Before moving to the gravitational description we would like to manipulate \eqref{schact} a bit to explain why a diffeomorphic invariant theory is necessary here. The Schwarzian derivative \( \sch{\phi}{\tau} \) enjoys a very interesting property that allows for the exchange of base and target space in this theory. Namely:
\begin{align}
  \sch{\phi}{\tau} = -  \qty(\partial_\tau \phi)^2 \sch{\tau}{\phi}~.
\end{align}
Therefore, an alternative formulation of the Schwarzian theory is given by:
\begin{align}\label{schact2}
  \act_\alpha[\tau] = \ell \int_0^1 \frac{\dd{\phi}}{\beta \qty(\partial_\phi \tau)} \qty[ \sch{\tau}{\phi} + 2 \pi^2 \alpha] + \lambda \,\beta  \int_0^1 \dd{\phi} \, \partial_\phi \tau ~,
\end{align}Something interesting has happened. Our target symmetries have become base space symmetries associated to changes of coordinate \(\phi \mapsto \sigma(\phi)\) belonging to the relevant group. If we want to gauge them, we must promote the action above to one enjoying full diffeomorphism invariance. We consider such a theory in the next section.

\section{One-dimensional Quantum Gravity}
\label{sec:qg}

Let us follow the procedure outlined before to obtain the parent gauge theory from the Goldstone theory described above. We consider in this section the \(\alpha>-1\) case.

First, it is useful to repackage the degree of freedom \(\beta \partial_\phi \tau\) into a one form \(e_\phi\) which we will take to be our einbein. There is a diffeomorphic invariant modulus which we label
\begin{align}
  \beta \coloneqq \int_0^1 \dd{\phi} \, e_\phi~,
\end{align}
with $\beta \ge 0$. The Schwarzian action involves derivatives of this object. As \(e_\phi\) is a one form, these are not diffeomorphic covariant. We need to introduce a connection \(\Gamma_\phi\) to properly define covariant derivatives. Under a change of coordinates on the base space circle \(\phi \mapsto \sigma(\phi)\) we have:
\begin{align}
  e_\phi \mapsto e_\sigma           & = \qty(\partial_\phi \sigma )^{-1} e_\phi ~,                                                        \\
  \Gamma_\phi \mapsto \Gamma_\sigma & = \qty(\partial_\phi \sigma )^{-1} \Gamma_\phi +  \partial_\phi  \qty(\partial_\phi \sigma )^{-1}~.
\end{align}
This shows the existence of a second invariant which we denote
\begin{align}
  \gamma \coloneqq \int_0^1 \dd{\phi} \, \Gamma_\phi~.
\end{align}Notice that this invariant vanishes for the usual Christoffel connection. With these tools we can write a theory of one-dimensional gravity as follows:
\begin{align}
  \act_{\t{\textsc{qg}}}[e,\Gamma] \coloneqq \ell \int_0^1 \dd{\phi} \, e^\phi \qty[ \mathcal{D}_\phi  \qty(e^\phi\, \mathcal{D}_\phi\,  e_\phi) - \frac{1}{2} \qty(e^\phi\, \mathcal{D}_\phi\, e_\phi )^2 ]+ \lambda\int_0^1 \dd{\phi} \, e_\phi~, \label{qgact}
\end{align}
where
\begin{align}
  e^\phi = e_\phi^{-1}~,\qquad \mathcal{D}_\phi e_\phi = \partial_\phi e_\phi - \Gamma_\phi e_\phi~.
\end{align}
In order to write a partition function for this theory we need to define diffeomorphic invariant quantum measures. They are straightforward to write from the invariant norm of fluctuations as:
\begin{align}
  \norm{\var e_\phi}^2 = \int_0^1 \dd{\phi} \, e^{\phi} \qty(\var e_\phi)^2           & \quad \implies \quad \DD{e} = \prod_\phi \, \sqrt{e^\phi} \, \dd{e_\phi(\phi)}\label{mese}~, \\
  \norm{\var \Gamma_\phi}^2 = \int_0^1 \dd{\phi} \, e^{\phi} \qty(\var \Gamma_\phi)^2 & \quad\implies\quad \DD{\Gamma} = \prod_\phi \, \sqrt{e^\phi} \, \dd{\Gamma_\phi(\phi)}~.
\end{align}Our partition function is thus
\begin{align}\label{ZQG}
  \parti_{\t{\textsc{qg}}} = \int \frac{\DD{e} \, \DD{\Gamma}}{\vol(\diffs)}\ \ex{-\act_{\t{\textsc{qg}}}[e,\Gamma]}~.
\end{align}Let us connect this result with the computation of \(\parti_{\diffu}\). We can always gauge fix the connection down to a constant given by \(\gamma\) by picking the appropriate diffeomorphism. A constant shift of the coordinate \(\phi\) remains a compact \(\U(1)\) zero mode we must still divide by. Furthermore we separate the integral over the invariant \(\beta\) from the rest of \(e_\phi\) as:
\begin{align}
  e_\phi = \beta \,\tilde{e}_\phi, \qq{where} \int_0^1 \dd{\phi} \, \tilde{e}_\phi =1~.
\end{align}
We are left with
\begin{align}
  \parti_{\t{\textsc{qg}}} = \int \dd{\beta} \dd{\gamma} \, w(\beta,\gamma)  \int \frac{\DD{\tilde{e}}}{\vol(\U(1))}\ \ex{-\act^\t{gauge fixed}_{\t{\textsc{qg}}}[\tilde{e},\beta,\gamma]}~, \,
\end{align}
where:
\begin{align}
  \act^\t{gauge fixed}_{\t{\textsc{qg}}}[\tilde{e},\beta,\gamma] & = \frac{\ell}{\beta} \int_0^1 \frac{\dd{\phi}}{\tilde{e}_\phi} \qty[ \partial_\phi^2 \log \tilde{e}_\phi - \frac{1}{2}\qty( \partial_\phi \log \tilde{e}_\phi)^2+ \frac{1}{2} \gamma^2  ] + \lambda \beta~.
\end{align}
The gauged fixed action agrees with the Goldstone action making the identifications \(\tilde{e}_\phi= \partial_\phi \tau\) and \(\gamma= 2\pi \sqrt{\alpha}\).
Notice that the measure for the \(\tilde{e}_\phi\) integral is in principle complicated by the explicit appearance of this variable in the original \(\DD{e}\) and \(\DD{\Gamma}\) measures. Furthermore, the gauge-invariant variables \(\beta\) and \(\gamma\) appear directly in the evaluation of the zero mode volume integrals. These facts complicate the original Goldstone formulation of the theory.

In order to obtain the Goldstone action from Quantum Gravity we must introduce delta functions that localize the moduli to specific values. In this case we must insert:
\begin{align}\label{diffu11}
  \parti_{\diffu}[\beta, \alpha] = \ev{\delta\qty[ \frac{1}{\beta}\qty(\int_0^1 \dd{\phi}\, e_\phi - \beta)]\; \delta\qty[\int_0^1\dd{\phi}\, \Gamma_\phi - 2 \pi \sqrt{\alpha}] }_{\t{\textsc{qg}}}.
\end{align}
Notice that the extra factor of \(\beta\) inside the first delta function is a consequence of the fact that in the original variables we need to insert:
\begin{align}
  \delta\qty[ \int_0^1 \dd{\phi} \, \partial_\phi \tau - 1 ] =\delta\qty[ \frac{1}{\beta}\int_0^1 \dd{\phi} \, e_\phi - 1 ] = \beta\, \delta\qty[ \int_0^1 \dd{\phi} e_\phi - \beta ]~.
\end{align}The computation of this path integral is simpler by following the alternative path of gauge fixing \(e_\phi\), as was the case in the simple \(\U(1)\) example discussed in section \ref{sec:u1}. This is a standard calculation but we will do it explicitly. The measure for diffeomorphisms \(\sigma(\phi)\) can be written explicitly as:
\begin{align}
  \norm{ \var \sigma}^2 = \int_0^1 \dd{\phi} \, e_{\phi}^3\, \qty(\var \sigma)^2  \, \quad\implies\quad \, \DD{\sigma} = \prod_\phi \, e_\phi^{\frac{3}{2}} \, \dd{\sigma(\phi)}~.
\end{align}
Any \(e_\phi\) can be written by gauge fixing it to a constant \(\beta\) and then performing a  as:
\begin{align}
  e_\phi = \beta\, \partial_\phi \sigma.
\end{align}
From the measure over \(e_\phi\) \eqref{mese} around our gauge fixed values \(\beta\) and \(\sigma=\phi\) we obtain:
\begin{align}
  \norm{ \var e_\phi}^2 = \frac{\var \beta^2}{\beta} + \int_0^1 \dd{\phi} \, \beta\, \qty(\partial_\phi \var \sigma)^2  \, \quad\implies\quad \DD{e} = \frac{\dd{\beta}}{\beta} \prod_{n \neq 0} \dd{\sigma_n}.
\end{align}
In the above we have written the path integral over Fourier modes on the circle, labeled by \(n\). In the Jacobian there was a \(\beta\) independent functional determinant which we have discarded and have used zeta function regularization to collect factors of \(\beta\). The zero mode left is compact and of unit size, so dividing by it is trivial. We are left with:
\begin{align} \nonumber
  \parti_{\diffu}\qty[\bar \beta, \alpha] & = \int_0^\infty \frac{\dd{\beta}}{\beta} \dd{\gamma}\,  \ex{-\frac{\ell \gamma^2}{2\beta} - \lambda \beta} \,  \beta\,\delta\qty[ \beta - \bar \beta ]  \;\delta\qty\big[\gamma - 2 \pi \sqrt{\alpha}] \prod_{n \neq 0}\int  \dd{\Gamma_n} \exp(- \frac{\ell}{2\beta} \sum_n \Gamma_n^2) \quad \quad   \\
                                          & = \qty(\frac{\ell}{\bar{\beta}})^{\frac{1}{2}}  \exp(-\frac{2 \pi^2 \ell \alpha}{\bar \beta} - \lambda \bar \beta)~. \label{diffu1res}
\end{align}This agrees with the result (\ref{u1sch}) quoted in the previous section. Notice that the odd factor of \(\beta\) in front of the delta function in \eqref{diffu11} becomes natural in this calculation as it constructs the delta function with respect to the quantum gravity measure \(\dd{\beta} \, w(\beta) =\frac{\dd{\beta}}{\beta}\).

Interestingly, the result \eqref{diffu1res} actually holds as well for imaginary values of \(\gamma\) when \(\alpha \in (-1,0]\) which coincides with the full range of \(\alpha\) for \(\diffu\) theories \cite{Stanford:2017thb}. Since we are fixing the holonomy externally we are allowed this freedom. At \(\alpha=-1\) the structure of zero modes changes and the analytic continuation is not possible anymore. We will come to this case momentarily.

We have been calling this a theory of quantum gravity, but it might be unfamiliar to some readers to deal with a theory with an independent connection \(\Gamma\). We can massage a bit the action to turn it into something familiar.

First integrate in a new einbein \(g_\phi\) such that \(\Gamma\) is its associated connection \(\Gamma= \partial_\phi \log g_\phi\). Notice that this introduces a new unphysical symmetry related to rescalings of \(g_\phi\) which we must gauge. This is our familiar Weyl symmetry present in String Theory. In this case, as opposed to String Theory, we must include a gauge field \(a_\phi\) to remove the redundancy. Finally, in order to avoid dealing with two different einbeins we define a matter field by \(e_\phi = \frac{g_\phi}{X^2}\).

The action \eqref{qgact} becomes, after discarding the total derivative term:
\begin{align}\label{QGactX}
  \act_{\t{\textsc{qg}}}[X,g,a] = 2 \ell \int_0^1 \frac{\dd{\phi}}{g_\phi} \qty( \partial_\phi X - \frac{a_\phi}{2} X )^2 + \lambda \int_0^1 \dd{\phi} \frac{g_\phi}{X^2}~.
\end{align}
This is nothing else than a gauged version of Conformal Quantum Mechanics \cite{deAlfaro:1976vlx}. Here we see why we needed to include the gauge field. The matter field transforms under the new Weyl symmetry as:
\begin{align}
  X \mapsto \ex{\frac{\omega}{2}} X~, \qquad g_\phi \mapsto \ex{\omega} g_\phi~, \qquad a_\phi \mapsto a_\phi + \partial_\phi \omega~.
\end{align}In critical String Theory, matter fields are  neutral under Weyl and therefore a connection is not needed to gauge the symmetry.  Our path integral has become:
\begin{align}
  \parti_{\t{\textsc{qg}}} = \int \frac{\DD{X} \DD{g} \DD{a}}{\vol(\diffs \times \t{Weyl})}\ \ex{-\act_{\t{\textsc{qg}}}[X,g,a]}~.
\end{align}with uniquely fixed local and gauge invariant measures:
\begin{align}
  \DD{X} & = \prod_\phi \, \sqrt{g_\phi} \; \frac{\dd{X(\phi)}}{X(\phi)^2}~,   \\
  \DD{g} & = \prod_\phi \, \sqrt{g^\phi} \; \frac{\dd{g_\phi(\phi)}}{X(\phi)}~, \\
  \DD{a} & = \prod_\phi \, \sqrt{g^\phi} \, X(\phi) \, \dd{a_\phi(\phi)}~.
\end{align}
Notice these are not the standard flat measures from quantum mechanics. The above definition is necessary to preserve the gauge symmetries and keep the partition function identical to the one defined in terms of \(e_\phi\) and \(\Gamma_\phi\). The range of \(X(\phi)\) is over \(\mathbb{R}_{>0}\). In this formulation the invariants are given by the two natural holonomies:
\begin{align}
  \beta \coloneqq \int_0^1 \dd{\phi} \frac{e_\phi}{X^2} \qquad\qq{and}\qquad \gamma \coloneqq \int_0^1 \dd{\phi} \, a_\phi~.
\end{align}Needless to say, the partition function of this theory agrees manifestly with \eqref{diffu1res} if we fix the values of the holonomies.

This is the closest thing to String Theory one could hope for in one dimension. While we have obtained this theory here in order to reproduce, by fixing the holonomies, the Goldstone theory we denoted \(\diffu\) we should consider it more generally as a bona fide theory of Quantum Gravity. We return to this question in section \ref{sec:qg1}.

Now we move on to the special case \(\alpha=-1\).

\section{The Schwarzian and Conformal Quantum Gravity}
\label{sec:cqg}

The Goldstone theory given by the parameter \(\alpha=-1\) is of a special character. At this particular value the symmetry is enhanced to a full non-compact \(\SL(2,\mathbb{R})\). As there exist 3 zero modes, we expect to require a minimum of 3 gauge fields to describe this action as opposed to the single connection \(\Gamma_\phi\) appearing in \eqref{qgact}. One simple way to guess the appropriate action is to exploit the connection to conformal quantum mechanics uncovered above. We could consider the matter Lagrangian:
\begin{align}
  \act_\t{\textsc{cqm}}[X] = 2 \ell \int_0^1 \dd{\phi} \qty[\qty( \partial_\phi X )^2  - \pi^2 X^2]+ \int_0^1 \dd{\phi} \frac{\lambda}{X^2}~.
\end{align}
One might think that the addition of the (negative) mass term breaks the conformal symmetry, but this is not the case as \(X^2\) is itself one of the symmetry generators of the group. The result is that, in this theory, \(\phi\) evolution represents the action of a compact symmetry generator, as opposed to the case where the mass term is absent \cite{deAlfaro:1976vlx,Bagrets:2016cdf}. All we need is to couple this theory to gauge fields for the \(\SL(2,\mathbb{R})\) currents and make them dynamical. The result is a theory of conformal quantum gravity  \cite{Marnelius:1978fs,Bars:1998ph}:
\begin{align}
  \act_\t{\textsc{cqg}}[X,g,a,b] =2 \ell \int_0^1 \frac{\dd{\phi}}{g_\phi} \qty[\qty( \partial_\phi X - \frac{a_\phi}{2} X )^2 + \qty(b_{\phi\phi} - \pi^2 g_\phi^2 ) X^2  ]+\lambda \int_0^1 \dd{\phi} \frac{g_\phi}{X^2}~.
\end{align}
where we have introduced an einbein \(g_\phi\), a gauge field \(a_\phi\) and a tensor \(b_{\phi \phi}\). This action is invariant under the following finite gauge transformations:
\begin{alignat}{4}
  \phi               & \longmapsto & \sigma\ \ \                  & = \sigma(\phi)                                                                                                                                                               \\
  X(\phi)            & \longmapsto & X'(\sigma)                   & = \ex{\frac{\omega(\phi)}{2}}\, X(\phi)                                                                                                                                      \\
  g_\phi(\phi)       & \longmapsto & g'_{\sigma} (\sigma)         & = \qty(\partial_\phi \sigma)^{-1} \ex{\omega(\phi)} g_\phi(\phi)                                                                                                             \\
  a_\phi(\phi)       & \longmapsto & a'_\sigma(\sigma)            & =\qty(\partial_\phi \sigma)^{-1} \qty[a_\phi(\phi) + \partial_\phi \omega(\phi) + g_\phi\, \kappa(\phi) ]\quad\quad                                                          \\
  b_{\phi\phi}(\phi) & \longmapsto & \  b'_{\sigma\sigma}(\sigma) & = \qty(\partial_\phi \sigma)^{-2} \left[ b_{\phi\phi}(\phi) + \pi^2 g_\phi^2 \qty(\ex{2\omega(\phi)} -1) - \Half g_\phi(\phi)\, \partial_\phi \kappa(\phi)-\right. \nonumber \\ & &&\phantom{=\ }- \left.\Half a_\phi(\phi)\, g_\phi(\phi)\, \kappa(\phi) - \frac{1}{4} g_\phi(\phi)^2 \kappa(\phi)^2  \right]~,
\end{alignat}
where \(\sigma(\phi)\) is associated to diffeomorphisms, \(\omega(\phi)\) is related to Weyl transformations and \(\kappa(\phi)\) is a new parameter for a gauged conformal symmetry.
With these fields we can define the following partition function:
\begin{align}\label{CQGpart}
  \parti_\t{\textsc{cqg}} = \int \frac{\DD{X} \DD{g} \DD{a} \DD{b}}{\vol(\diffs \times \t{Weyl}\times \t{conf})}\ \ex{-\act_\t{\textsc{cqg}}[X,g,a,b]}~.
\end{align}The uniquely gauge invariant quantum measures are given by
\begin{align}
  \DD{X} & = \prod_\phi \, \sqrt{g_\phi} \; \frac{\dd{X(\phi)}}{X(\phi)^2}                        \\
  \DD{g} & = \prod_\phi \, \sqrt{g^\phi} \; \frac{\dd{g_\phi(\phi)}}{X(\phi)}                     \\
  \DD{a} & = \prod_\phi \, \sqrt{g^\phi} \, X(\phi)\, \dd{a_\phi(\phi)}                           \\
  \DD{b} & = \prod_\phi \, \qty(g^\phi)^{\nicefrac{3}{2}} \, X(\phi)^3\, \dd{b_{\phi\phi}(\phi)}~.
\end{align}As a sanity check, one can show that the gauged fixed version of this action on a trivial background reproduces the Schwarzian action. If we consider the trivial gauge configuration
\begin{equation}
  g_\phi=1~, \quad a_\phi=0~, \qq{and} b_{\phi\phi}=0~,
\end{equation}
and relabel \(X(\phi) = \qty(\beta \partial_\phi \tau)^{-\frac{1}{2}}\) we recover \eqref{schact2}. Notice that this is a correct parameterization of the \(X\) field as the following holonomy is gauge invariant in this theory and must be fixed to recover the Schwarzian partition function:
\begin{align}
  \beta = \int_0^1 \dd{\phi} \frac{g_\phi}{X^2}~.
\end{align}
What about the zero modes? They are also exactly reproduced. The trivial gauge configuration is invariant under transformations satisfying:
\begin{align}
  \omega & = \log \partial_\phi\sigma~,                                            \\
  \kappa & = -\partial_\phi\omega~,                                                \\
  0      & = 2\pi^2 \qty[ \qty(\partial_\phi \sigma)^2 -1] + \sch{\sigma}{\phi}~.
\end{align}
The solutions to the above equations are precisely the global \(\SL(2,\mathbb{R})\) modes that we are gauging.

Let us now turn to the computation of the partition function \eqref{CQGpart}. We will try to follow the same path as before. We perform a gauge fixing where we can use all gauge transformations and leave no zero modes behind. Then the only non-trivial part comes in the fixing of the gauge invariant quantities to the trivial sector of the theory. We fix:
\begin{align}
  X = \beta^{-\frac{1}{2}}~,\qquad g_\phi =1~, \qquad a_\phi=0~.
\end{align}
We are left with:
\begin{align}\label{ZCQG1}
  \parti_\t{\textsc{cqg}} = \int_0^\infty \frac{\dd{\beta}}{\beta}\ \ex{\frac{2 \pi^2 \ell}{\beta} - \lambda \beta} \int \DD{b}\ \exp(-2 \frac{\ell}{\beta} \int_0^1 \dd{\phi} \, b_{\phi\phi})~.
\end{align}
Here we see an interesting peculiarity. The functional integral over \(b_{\phi\phi}\) is badly divergent. It is true that we have not specified a region of integration, but under the naive guess that it runs over all real functions we would encounter an infinite result. Higher derivative terms could regulate this behavior and we will come back to this issue in the next section. For now, what interests us is the application of this theory in the computation of \(\parti_\t{Schwarzian}\). The integral over \(\beta\) is easy to deal with as we did in the \(\diffu\) case before. We just insert the appropiate \(\beta \, \delta (\beta - \bar \beta)\) factor in the path integral. What would be next is the insertion of three more delta functions corresponding to the moduli associated to the gauge fields \(g_\phi, a_\phi\) and \(b_{\phi \phi}\). The problem here is that, while these gauge invariant quantities exist, they are not given by integrals of local expressions. Luckily, we do not need to know these exact expressions. We will be inserting them inside delta functions. As such we only need to know these quantities very near the manifold of field configurations gauge equivalent to the trivial background. So for each field \(b_{\phi \phi}\) which is related to 0 by a gauge transformation, we need to know these expressions. Let us first consider doing this around \(b_{\phi\phi}=0\). In the vicinity of this point all trivial field deformations are generated by:
\begin{align}\label{bgauge}
  \var b^{\t{pure gauge}}_{\phi\phi} = \frac{1}{2} \qty[4\pi^2 \partial_\phi \xi + \partial_\phi^3 \xi ]~, \qq{with} \sigma = \phi + \xi~.
\end{align}
Setting \(\var b^{\t{pure gauge}}_{\phi\phi}=0\) yields 3 independent (normalized) solutions:
\begin{align}
  \xi^{-1} = \sqrt{2} \cos(2 \pi \phi)~, \quad \xi^{0} = 1~, \quad \xi^{1} = \sqrt{2} \sin(2 \pi \phi)~.
\end{align}
It is a simple exercise to check that the following quantities are invariant under variations of the field \(b_{\phi\phi}\) of the form \eqref{bgauge} :
\begin{align}
  \var \mu^I = \ell \int_0^1 \dd{\phi} \, \frac{\var b_{\phi\phi}}{\bar \beta} \, \xi^I, \qq{with} I \in\set{-1, 0, 1}~.
\end{align}
The \(\bar \beta\) dependence of this term is dictated by diffeomorphism invariance, as \(\var b_{\phi\phi}\) transforms as a weight two tensor. We have also added an overall factor of \(\ell\) for dimensional analysis.

Therefore, these constitute the first order variations of the moduli near the manifold of pure gauge configurations, around \(b_{\phi\phi}=0\). We just need to insert delta functions for these objects in our partition function. This procedure must be performed locally around each point in the manifold of pure gauge configurations. We must then integrate over it by patching the result computed in the tangent space of each point. This procedure is analogous to the patch theory of the Fermi Liquid state \cite{Polchinski:1992ed}. In order to capture the curvature of this manifold we must go beyond leading order to obtain the next order corrections that regulate the tangent space result which is infinite due to target space divergences. Instead of doing this exactly we can resort to effective theory and just add to the action a contribution from the first analytic correction in powers of \(\var b_{\phi \phi}\), thus capturing the curvature of the trivial configuration manifold. While the Fermi energy provides the relevant scale in the Fermi Liquid, here the only scale is \(\ell\). Therefore, we can add a quadratic correction to the action resulting in a path integral of the patch theory:
\begin{align}
  \parti_{\t{Schwarzian}}^{\t{patch}} & = \ex{\frac{2 \pi^2 \ell}{\bar \beta} - \lambda \bar\beta} \int \DD{\var{b}}\ \prod_I \delta\qty[\var \mu^I ] \exp(-2\frac{\ell}{\bar\beta} \int_0^1 \dd{\phi} \var b_{\phi\phi} -  \frac{\ell^3}{\Delta B^2\bar \beta^3} \int_0^1 \dd{\phi} \var b_{\phi \phi}^2)~,\label{patch1}
\end{align}
where \(\Delta B\) is the typical size of the patch. The \(\bar \beta\) dependence of the expression above can again be obtained by noticing that diffeomorphism invariance demands it. We will write a full gauge invariant version of this type of term in the next section when we discuss applications to quantum gravity.

We are now ready to perform this reduced path integral. We can integrate directly the three zero modes against the delta functions. This spits out factors of \(\frac{\bar \beta}{\ell}\).
\begin{align}
  \parti_{\t{Schwarzian}}^{\t{patch}}\qty[\bar\beta]
   & =  \ex{\frac{2 \pi^2 \ell}{\bar \beta} - \lambda \bar\beta} \qty(\frac{\bar \beta}{\ell})^3 \prod_{n \neq I} \int \dd{\var b_n} \exp( - \frac{\ell^3}{\Delta B^2\bar \beta^3} \sum_{n \neq I} \, \var b_{n}^2) \nonumber \\
   & =\qty(\prod_{n \neq I} \Delta B ) \ \ex{\frac{2 \pi^2 \ell}{\bar \beta} - \lambda \bar \beta} \qty(\frac{\ell}{\bar\beta})^{\frac{9}{2}-3} \nonumber                                                                     \\
   & =  \vol(\t{patch})\ \qty(\frac{\ell}{\bar\beta})^{\frac{3}{2}}\ \ex{\frac{2 \pi^2 \ell}{\bar\beta} - \lambda \bar\beta}~.
\end{align}
In the above expression we have used as usual zeta functional regularization to collect the infinite number of \(\frac{\ell}{\beta}\) factors, accounting for the zero modes. We have also subtracted the inconsequential overall normalization.

Now, we could repeat this procedure over different points of the manifold of trivial configurations. Gauge invariance, however, implies the same \(\bar\beta\) dependence. Up to a volume of this manifold (which should be compact for  \(\parti_\t{Schwarzian}\) to be well defined) we can quote the final result:
\begin{align}
  \parti_\t{Schwarzian}\qty[\bar \beta] =\sum_{\t{patches}}\parti_\t{Schwarzian}^{\t{patch}}\qty[\bar\beta] =\qty(\frac{\ell}{\bar \beta})^{\frac{3}{2}}\ \ex{\frac{2 \pi^2 \ell}{\bar \beta} - \lambda \bar \beta}~.
\end{align}This agrees with \eqref{stanwit}. Notice that this result was obtained from a manifestly local theory with explicitly defined local quantum measures. All non-localities came from the freezing of the gauge invariant moduli. Although some technology was required, the computation followed standard path integral methods without resorting to localization or Hamiltonian arguments.

The arguments here establish a concrete formulation of the Schwarzian theory in terms of a theory of (Conformal) Quantum Gravity.

\section{UV finiteness in one-dimensional Quantum Gravity}
\label{sec:qg1}

One of the most ambitious open problems in physics is the formulation of a manifestly UV finite theory of Quantum Gravity. String Theory proposes a perturbative solution to this problem by producing quantum geometry in target space. Although an explicit proof of its UV finiteness in its loop expansion is absent, it is expected to hold true. The price to pay, in critical String Theory, is a large dimension \(d= \{10, 26\}\) of the space-time and an infinite tower of fields. In order to have a more tractable problem, much attention has been devoted to the study of Quantum Gravity in low space-time dimensions. At least for \(d \leq 3\), where metric fluctuations become non-dynamical, we have some hope that a theory of gravity (and maybe a few matter fields) can be UV finite. So far, this program has not produced a fully understood working example in \(d=3\). What about even lower dimensions? In \(d=2\) obvious candidates are the worldsheet theory of critical String Theory and Liouville Gravity (emergent in the non-critical case). Let us look at each of these cases in detail.

Let us start with critical String Theory. While there has been recent progress in the understanding of the sphere and disc partition functions \cite{Anninos:2021ene,Eberhardt:2021ynh,Erbin:2019uiz}, let us concentrate on the long understood torus topology. The partition function in this topology is well known. In the bosonic case, it is divergent. While modular invariance indicates that this divergence can always be thought as arising from the IR behavior of the theory (i.e. the tachyon mode), it is still a fact that this theory cannot be considered a complete theory of two-dimensional Quantum Gravity. This situation is, of course, corrected for Superstrings. The price to pay is high. The torus partition function becomes identically zero, see for example \cite{Polchinski:1998rr}, and cannot be used as a basis for the statistical mechanics of Euclidean geometries. Even if one is willing to accept this, notice that, from the worldsheet perspective, the way this solution comes about is unusual. Different sectors of the theory enter the partition function with different signs giving rise to a cancellation of the divergence (and of everything else in this case.). While we are used to renormalizing partition functions multiplicatively, additive subtractions are not typically allowed within the standard paradigm. From a worldsheet Hilbert space perspective, the subtractions correspond to the removal of states (graded by the fermion number) from the spectrum -- a phenomenon we will refer to as addition of \emph{ghost} states. For Superstring Theory this cancellation is complete.

Liouville theory is another well studied example of \(d=2\) Quantum Gravity. In this case, the partition function shows divergences for geometries with very small areas \cite{Seiberg:1990eb,Zamolodchikov:1982vx,Muhlmann:2021clm}.% SAY MORE?

Finally, let us comment on the simplest possible example: one-dimensional Quantum Gravity. Quite generally these theories can be (formally) formulated in the following form:
\begin{align}
  \parti_{1d} = \int_0^\infty \frac{\dd{\beta}}{\beta} \parti_\t{matter}[\beta] \qq{with} \parti_\t{matter} = \Tr \ex{-\beta H}~,
\end{align}
where \(\parti_\t{matter}\) refers to the partition function of some quantum mechanical system with Hamiltonian \(H\). The measure of integration over inverse temperatures \(\beta\) is standard  \cite{Polyakov:1987ez} and arises from the translational zero mode of diffeomorphisms of the circle, as we saw in a few examples above.

The point here is that \(\parti_{1d}\) is always divergent provided \(H\) is the Hamiltonian of a well behaved quantum system. Whenever the dimension of the Hilbert space is infinite \(\parti_\t{matter} \to \infty\) as \(\beta \to 0\). Even if the dimension of the Hilbert space,  \(\dim \mathcal{H}\),  is finite we have:
\begin{align}\label{QGdiv}
  \parti_{1d}  \to \dim \mathcal{H}\; \log \Lambda~, \qq{as} \beta \to 0~,
\end{align}
where \(\Lambda\) is a high energy cut-off.\footnote{An analogous divergence is present in 2d quantum gravity on the torus, and discussed in this context in \cite{Bershadsky:1990xb,Anninos:2020geh}.}

These models of Quantum Gravity have a natural target space interpretation as the worldline formalism for first quantized field theory. But the connection between these formalisms matches the Quantum gravity partition function to the one-loop effective action as:
\begin{align}
  \parti_{1d} \sim \log \parti^\t{1-loop}_\t{\textsc{qft}}~.
\end{align}In this setup the UV divergences discussed above are naturally reinterpreted as 1-loop QFT divergences. In this framework, it is natural to have additive cancellations of these infinities as this is the standard technology for the 1-loop effective action. From the worldline perspective, however, the situation is slightly embarrassing. This is the simplest situation one could hope for and we have yet to produce a fully UV-complete theory of Quantum Gravity.

After this long digression, let us take a look at the models we discussed in previous sections and consider them as theories of Quantum Gravity. By this, we mean that we will be integrating over the modulus \(\beta\).

Let us start with the theory defined by \(\parti_{\t{\textsc{qg}}}\) in \eqref{ZQG} either parameterized by the fields \(e_\phi\) and \(\Gamma_\phi\) or \(X\), \(g_\phi\) and \(a_\phi\), as it makes no difference. If we insert no delta function to fix the holonomies we obtain:
\begin{align}\label{QGZ2}
  \parti_{\t{\textsc{qg}}} = \int \frac{ \dd{\beta} \, \dd{\gamma} \, \ell^{\frac{1}{2}}}{\beta^{\frac{3}{2}}}\ \ex{-\frac{\ell \gamma^2}{2\beta} - \lambda \beta}~.
\end{align}Let us, for now keep the holonomy \(\gamma\) fixed and perform the \(\beta\) integral. The result is:
\begin{align}\label{QGgamma}
  \parti_{\t{\textsc{qg}}} = \frac{\ex{-|\gamma| \sqrt{\frac{\lambda \ell}{2}}}}{|\gamma|}~.
\end{align}
This is manifestly finite. How could this happen, given the previous discussion? In this theory the Weyl holonomy  \(\gamma\) is fixed. This is a non-local procedure that removes the interpretation of the matter content of the theory as that of a normal quantum mechanical system with a well defined Hilbert space. If one looks at the statistical distribution of inverse temperatures \(\beta\) in \eqref{QGZ2} one can see something interesting. At low temperatures \(\beta \gg \gamma^2 \ell\), the system behaves like a simple one-dimensional particle with a density of state going like \((E-\lambda)^{-\frac{1}{2}}\) above a gap \(\lambda\) coupled to Quantum Gravity. While this looks harmless, it is important to remark that this density of states persists all the way down to \(E \to \lambda\) so it is not the standard density of states per unit volume valid at energy scales larger than the confining box of the system. Still, all states contribute positively to the partition function. But at inverse temperatures of order \(\beta \sim \gamma^2 \ell\) something happens. The distribution in \(\beta\) drops exponentially fast to zero and one-dimensional geometry ceases to exist. This is exactly the type of behavior one could expect of a theory of Quantum Gravity at the Planck scale \(\lp \coloneqq \gamma^2 \ell\). From this perspective, \(\ell\) seems more like the string scale. In any case, it is the expectation value of the holonomy that gives rise to an emergent \(\lp\). Therefore, the end of geometry in the UV is an emergent phenomenon in this case.

From a Holographic perspective, the theory \eqref{QGgamma} is dual to the theory of curves near a single boundary of the trumpet geometry of \(\ads_2\) where the length \(\beta\) of the boundary curve is allowed to fluctuate (see for example \cite{Anninos:2017hhn,Stanford:2020qhm,Gross:2019ach}). Small sizes are seemingly regulated by a geometric feature in the deep interior. As such, coupling the boundary quantum mechanics to dynamical Quantum Gravity disrupts the traditional UV/IR correspondence \cite{Susskind:1998dq} expected to hold in \ads/\cft.

Finally, from the worldline Hilbert space perspective, it is inevitable that this depletion in the inverse temperature distribution originates in the appearance of \emph{ghost} states at the Planck length scale. This can be checked explicitly by noticing that the density of states of this quantum system is
\begin{align}
  \rho(E) \sim \ell^{\frac{1}{2}}\; \frac{\cos(\sqrt{2 \gamma^2 \ell (E-\lambda)})}{(E-\lambda)^{\frac{1}{2}}}~.
\end{align}
While at low energies, \(E \sim \lambda\), we reproduce the expected square root behavior discussed above, at high energies \(E \sim \frac{1}{\lp}\) \emph{ghost} states enter the spectrum. This is not a bug of this theory but a failure of arriving at a Hilbert space interpretation. The Euclidean path integral for this system is well posed, positive and finite. Given the discussion above, one should think of the breakdown of the Hilbert space as a feature and not a problem in any UV-complete Quantum Gravity. As a last check of our ideas above one could compute the total number of states in the theory up to a given energy \(E\) and call it \(\dim \mathcal{H}(E)\). We easily obtain:
\begin{align}
  \dim \mathcal{H}(E) = \int_\lambda^E  \dd{E'} \rho(E')  \sim \frac{\sin(\sqrt{2 \gamma^2 \ell (E-\lambda)})}{\abs{\gamma}}~,
\end{align}
which can be seen to oscillate around \(0\). This is necessary of any UV finite theory of Quantum Gravity to remove the divergence in \eqref{QGdiv}.

In this theory, all (reasonable) observables are well defined and one could compute correlation functions of \(\beta\) or of matter fields \(X\) provided they are diffeomorphism invariant. The situation is completely analogous to String Theory and warrants further study.

Let us now turn to Conformal Quantum Gravity given by the theory \eqref{CQGpart}. As we remarked, this path integral when integrating over all degrees of freedom, including holonomies, is divergent. There is a simple remedy for this. It turns out that there is an infinite collection of corrections to this action that are gauge invariant and given by higher derivative terms. Consider the first such term:
\begin{align}
  \Delta \act_\t{\textsc{cqg}} = \ell^3 \int_0^1 \dd{\phi}\frac{X^2}{g_\phi^3}  \qty[\qty( \partial_\phi X - \frac{a_\phi}{2} X )^2 + \qty(b_{\phi\phi} - \pi^2 g_\phi^2 ) X^2  ]^2~.
\end{align}
We could add this term to the action and consider \(\act_{q\t{\textsc{cqg}}} = \act_\t{\textsc{cqg}} + q \, \Delta \act_\t{\textsc{cqg}}\), where \(q\) is a positive dimensionless coupling constant. We can use the same techniques as before to compute the path integral obtaining a formula analogous to \eqref{ZCQG1}. We have
\begin{align}
  \parti_{q\t{\textsc{cqg}}} = \int_0^\infty \frac{\dd{\beta}}{\beta}\ \ex{ - \lambda \beta} \int \DD{b}\ \exp(-2 \frac{\ell}{\beta} \int_0^1 \dd{\phi} \, (b_{\phi\phi} -\pi^2)- q \frac{\ell^3}{\beta^3} \int_0^1 \dd{\phi} (b_{\phi\phi} - \pi^2)^2 )~.
\end{align}
Notice that this correction (for the full theory) is of a similar form to the correction induced in the patch theory by curvature effects \eqref{patch1}, including the \(\beta\) scaling. We can now easily perform the remaining gaussian integral to obtain:
\begin{align}
  \parti_{q\t{\textsc{cqg}}} = \int_0^\infty \frac{\dd{\beta}}{\beta}\ \ex{ - \qty(\lambda - \frac{1}{q \ell} ) \beta}~.
\end{align}
The result shares, of course, the same UV problems of any theory of Quantum Gravity once we set free all non-local gauge invariant observables. Looking at the above result we would interpret the matter content of Conformal Quantum Gravity as a single state of energy \(\lambda -\frac{1}{q\ell}\).  Here we clearly see what the role of \(q\) was in regulating the theory.

\section{Discussion}
\label{sec: Discussion}

In this note we discussed the relation between Goldstone and Gauge theories and gave a simple prescription to compute partition functions of the first kind in terms of observables in the second type of theory. This prescription turns the formulation of the problem and theory into a purely local form while non-localites are introduced by the particular observables computed. In particular, the recipe amounts to freezing all moduli of the theory to a desired value and computing this restricted partition function in the Gauge theory.

We have applied this formalism to the Schwarzian and \(\diffu\) Goldstone theories and explained why all physics in these theories is contained in theories of one-dimensional (Conformal) Quantum Gravity. Using this formalism we have provided alternative derivations of the partition functions in these theories. We expect this technology can be successfully used to compute and provide insights on the space of interesting observables.

Having presented these theories, we consider the problem of a defining a UV complete theory of Quantum Gravity in one dimension. We associate possible UV divergences to the dimension of the Hilbert space. Therefore, we argue that any finite theory cannot have a standard Hilbert space interpretation without the inclusion of \emph{ghost} states in its spectrum. We present examples in String Theory where this happens manifestly. The examples of Quantum Gravity we present can naturally exhibit this feature. It is the freezing out of some modulus that destroys the Hilbert space interpretation and introduces \emph{ghost} states. In the example presented, \emph{ghost} states only come in at the Planck scale. This natural length scale is emergent in the theory and set by the value of the frozen modulus. It is only at that scale that the Hilbert space interpretation breaks down. Therefore, at low energies there is a standard quantum mechanical picture which only disappears when base space geometry stops making sense. This seems like a perfect example of what one would expect for a UV complete theory of Quantum Gravity. It would be interesting to find more involved and phenomenologically viable examples in higher dimensions.

Lastly, let us comment on possible applications of these ideas to holography. While the Goldstone theories discussed have a well understood holographic dual in terms of near \(\ads_2\) geometries \cite{Maldacena:2016upp,Jensen:2016pah}, it would be interesting to further extend the dictionary to theories where (some) boundary moduli are set free and, therefore, we would have a duality between two gravitational theories as in \cite{Karch:2000ct,Compere:2008us}. This is particularly interesting in the context of \(\ds\) holography (see for example \cite{Alishahiha:2004md}) and in the dual understanding of bulk observers, where the worldlines in question necessarily fluctuate \cite{Anninos:2011af,Anninos:2018svg,Anninos:2017hhn}.

It is often speculated \cite{Gibbons:1977mu,Parikh:2004wh,Banks:2006rx,Witten:2001kn}, that quantum gravity on compact spatial slices, including those surrounded by a cosmological horizon, is described by an underlying finite dimensional Hilbert space. This might echo the interplay between the divergences exhibited by the theories of one-dimensional Quantum Gravity presented above (and their two-dimensional counterpart \cite{Anninos:2020geh,Kutasov:1990sv}) and the appearance of {\it{ghost}} states.
% We hope to have more to say about this connection in the near future.

\section*{Acknowledgements}
It is a pleasure to acknowledge interesting discussions with Frederik Denef, Beatrix Mühlmann, Damián Galante, and Eric Perlmutter. D.H. is supported in part by the ERC starting grant \textsc{gengeohol} (grant agreement N\textsuperscript{\underline{\scriptsize o}} 715656). D.A. is funded by the Royal Society under the grant The Atoms of a deSitter Universe. All authors would like to credit the Physics Sessions Initiative, where some of this work was conducted.

\bibliographystyle{jhep}
\bibliography{bibliographyCQM}
% \printbibliography

\end{document}